\documentclass[10pt,twocolumn,aip,jcp,superscriptaddress,notitlepage,longbibliography,nofootinbib]{revtex4-1}

\makeatletter
\let\@fnsymbol\@fnsymbol@latex
\@booleanfalse\altaffilletter@sw
\makeatother

\usepackage{amsmath}
\usepackage{graphicx}
\usepackage{color}
\usepackage{upgreek}
\usepackage{enumerate} 
\usepackage[linkcolor = blue, citecolor = blue, urlcolor = blue, colorlinks = true]{hyperref}
\usepackage[version=3]{mhchem}
\usepackage{commath,amssymb}
\usepackage{ushort}
\usepackage{multirow}
\usepackage[usenames,dvipsnames]{xcolor}
\usepackage{cleveref}
\usepackage{epstopdf}
\usepackage{subcaption}
\usepackage[exponent-product = \cdot]{siunitx}

\usepackage{hyperref}

\usepackage{xr-hyper}

\newcommand{\jcp}{\color{black}}
\newcommand{\tg}{\emph{T\textsubscript{g}}}

\crefname{equation}{Eq.}{Eqs.}
\crefname{figure}{Fig.}{Figs.}
\crefrangeformat{equation}{Eqs.~#3(#1)#4--#5(#2)#6}

\definecolor{myred}{RGB}{162, 20, 47}
\definecolor{myblue}{RGB}{0, 114, 189}
\definecolor{mygreen}{RGB}{119, 172, 48}
\definecolor{myyellow}{RGB}{237, 177, 32}

\makeatother
\begin{document}

	
	
	\title{Relaxation Dynamics of a Liquid in the Vicinity of an Attractive Surface: The Process of Escaping from the Surface}
	
	

	\author{Alireza F. Behbahani}
	\email{aforooza@uni-mainz.de}
	\affiliation{Institut f\"{u}r Physik, Johannes Gutenberg-Universit\"{a}t Mainz, Staudingerweg 7, D-55099 Mainz, Germany}
	\affiliation{Institute of Applied and Computational Mathematics, Foundation for Research and Technology - Hellas, Heraklion GR-71110, Greece}

	\author{Vagelis Harmandaris}
	\affiliation{Computation-based Science and Technology Research Center, The Cyprus Institute, Nicosia 2121, Cyprus}
	\affiliation{Institute of Applied and Computational Mathematics, Foundation for Research and Technology - Hellas, Heraklion GR-71110, Greece}
	\affiliation{Department of Mathematics and Applied Mathematics, University of Crete, Heraklion GR-71110, Greece}

	\begin{abstract}
		We analyze the displacements of the particles of a glass-forming molecular liquid perpendicular to a confining solid surface, using extensive molecular dynamics simulations with atomistic models.
		In the vicinity of an attractive surface, the liquid molecules are trapped.
		Transient localization of liquid molecules near the surface introduces a relaxation process, related to the escape of molecules from the surface, into the dynamics of the interfacial liquid layer.
		To describe this process, we analyze several dynamical observables of the confined liquid.
		The self-intermediate scattering function and the mean-squared displacement of the particles located in the interfacial layer are dominated by the process of escaping from the surface.
		This relaxation process is also associated with a strong heterogeneity in the mobility of the interfacial particles.
		The studied model liquid is hydrogenated methyl methacrylate. 
		For the confining wall, we consider different models, namely 
		a periodic single layer of graphene and a frozen amorphous configuration of the bulk liquid (frozen wall). 
		Near graphene, where the liquid molecules form a layered structure and adopt parallel-to-surface orientation, a clear separation between small-scale  movements of the molecules near the surface and the process of escaping from the surface is observed. This is reflected in the three-step relaxation of the interfacial layer. However, near the frozen wall, where 
		the liquid molecules do not have a preferential alignment, a clear three-step relaxation is not seen, even though the dynamical quantities are controlled by the process of escaping from the surface\footnote{This manuscript has been accepted for publication in \emph{The Journal of Chemical Physics}; \href{https://doi.org/10.1063/5.0231689}{https://doi.org/10.1063/5.0231689}}.
	\end{abstract}
	
	\maketitle
	
	\section{Introduction}
	
	The dynamics of confined glass-forming liquids has been extensively studied computationally~\cite{scheidler2004relaxation,torres2000molecular,smith2003structural, barrat2010molecular,kob2012non,zhang2018we,horstmann2022structural} and experimentally~\cite{arndt1997length, krutyeva2009neutron,papon2012glass,panagopoulou2020substrate,cheng2020correlation,reuhl2022confinement}. These studies are partly motivated by the technological importance of understanding the behavior of these materials.
	Furthermore, confined systems have been studied in order to gain insight into the mechanism of glassy dynamics and the nature of glass transition temperature, \tg~\cite{ediger2014dynamics,schweizer2019progress}.
	
	Some features of the dynamics of glass-forming liquids are usually discussed based on the so-called cage effect~\cite{gotze2009complex,binder2011glassy,pastore2014cage,helfferich2014continuous}. According to it, at short times, a particle of the liquid is trapped in a temporary cage formed by its neighboring particles. 
	At intermediate times, some of the particles escape from their cages while some of them are still restricted in the cages; this leads to the heterogeneity of particle displacements within the liquid which can be recognized from the peak of the non-Gaussian parameter for the particle displacements~\cite{kob1997dynamical,reichman2005mode,vorselaars2007non,peter2009md}.
	The cage effect leads to a two-step relaxation of the liquid; a short-time step that comes from fast movements within the cage and a long-time step that appears after cage breaking ($\alpha$-relaxation)~\cite{binder2011glassy}.
	The two-step  relaxation can be seen through the two-step decay of, for example, self-intermediate scattering function or through
	the two-step increase of mean-squared displacement with time~\cite{binder2011glassy,berthier2011theoretical}.
	
	{\jcp Despite the two-step relaxation of bulk liquids, previously, for the interfacial polymer layer (the layer in the vicinity of the surface), in a polybutadiene melt confined between graphite walls, a three-step decay of self-intermediate scattering function was observed, using molecular dynamics simulation~\cite{yelash2012three,solar2017relaxation}.
		The three-step relaxation was discussed as a polymer-specific phenomenon due to the desorption of the segments of the chains from the surface~\cite{yelash2012three}.
		In the current work, we study the relaxation dynamics of the interfacial layer in a low molecular weight confined liquid and show that the dynamics of the interfacial layer is also affected by the desorption process, or the process of escaping of the liquid molecules from the surface, as we call it here.
		More importantly, however, we provide an extensive analysis of this process, mainly by analyzing 
		various dynamical observables of the interfacial layer and discussing their correlations.
		It should be noted that many experimental and simulation studies have shown the development of the surface-induced gradients in the relaxation dynamics of liquids~\cite{scheidler2003relaxation,starr2016bound,carroll2017analyzing,zhang2017effects,mapesa2020interfacial,foroozani2021gradient}. Particularly, near an attractive surface, the relaxation dynamics is slower than in regions far from the surface or in bulk. 
		Because of this effect, the \emph{overall} relaxation functions of thin liquid films or polymer nanocomposites might exhibit a long tail or a three-step decay~\cite{scheidler2003relaxation,starr2016bound,carroll2017analyzing,zhang2017effects,mapesa2020interfacial,foroozani2021gradient}. In such cases, the relaxation time of a portion of the liquid that is close the surface is well separated from the relaxation time of the portion far from the surface and this leads to the appearance of two separated steps in the overall relaxation function of the system (in addition to short-time vibrational motions).
		We emphasize that such an effect is not the subject of the current study.
		Rather, we focus on the relaxation dynamics of the first interfacial layer and show that the presence of an attractive surface introduces an additional process in its relaxation dynamics.
		All analyses and discussions of the current work aim to describe this relaxation process.
		This work goes beyond the previous contributions in the following aspects:
		(i) We show that the dynamics of the interfacial liquid layer in the direction perpendicular to the confining surface is governed by the process of escaping of molecules from the surface.
		We discuss the signature of this process in different dynamical observables of the first interfacial layer.
		(ii) For the analysis of this relaxation process, we study the effects of temperature and surface/liquid attraction strength on the dynamics of the interfacial liquid.
		(iii)
		We consider two different model confining surfaces: one, near which the liquid forms a layered structure, and another, which is realized by freezing an equilibrated configuration of the liquid and therefore does not perturb the structure of the liquid.
		(iv) As an idealized case in which the particles are not trapped near the surface and the surface only imposes a geometric effect, we analytically calculate the dynamical observables of a Fickian liquid near a reflective wall and compare them with the simulation results.
	}
	
	\section{Model}
	\label{sec:model}	
	We present the results of 
	molecular dynamics simulations  
	of a low molecular weight liquid confined between solid surfaces. 
	The model liquid is hydrogenated methyl methacrylate (HMMA, C$_5$H$_{10}$O$_2$) which is similar to the chemical repeating unit of poly(methyl methacrylate), a well-known glass-forming polymer.
	The selection of HMMA was also motivated by our preliminary work on confined PMMA melts, in which we observed three-step relaxation of the interfacial layer (data are not shown).  
	For the confining surface, we consider different models. We mainly report the results of HMMA liquid confined between periodic single layers of pristine graphene.  
	Besides the simulation of HMMA/graphene using an available force field (and calculating HMMA/graphene interactions via a mixing rule),  we have performed some simulation runs with adjusted HMMA/graphene interactions, to investigate the effect of the surface/liquid interaction strength. 
	We have also performed some simulations where the confining surface is a frozen amorphous configuration of the bulk liquid.

\begin{figure}[!htb]
		\centering
		\begin{subfigure}{0.45\textwidth} 
			\includegraphics[width=\textwidth]{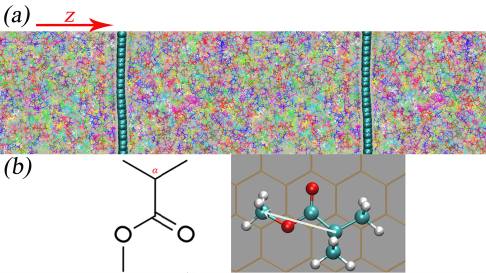}
		\end{subfigure}
		\begin{subfigure}{0.45\textwidth} 
			\includegraphics[width=\textwidth]{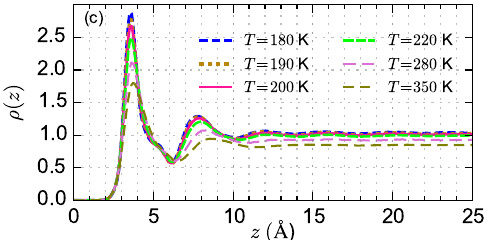}
		\end{subfigure}
		
		\caption{ (a) A snapshot of HMMA/graphene system. The simulation box and a part of one of its period images along the $z$ direction are shown. The thickness of the confined liquid film is around $10$ nm.
			(b)  A schematic representation of the chemical structure of HMMA ($\alpha$-carbon atom of HMMA is also shown) and a top view snapshot of a relaxed configuration of one HMMA molecule on the graphene surface.
			(c) Interfacial mass density profiles of the confined liquid as a function of distance from the confining surface.		}
		\label{Fig:model1} 
	\end{figure}

	Simulations were performed with atomistic models using the GROMACS package~\cite{pronk2013gromacs}.   
	Periodic boundary condition was applied in all directions.
	In the case of HMMA/graphene system, the dimensions of the simulation box are almost $5 \times 5 \times 10$ nm$^3$ along the $x$, $y$, and $z$ directions, respectively (the thickness of the confined liquid is around $10$ nm.).
	\autoref{Fig:model1}a shows a snapshot of the liquid confined between graphene surfaces.
	The leap-frog algorithm with $1$ fs time step was used for the integration of the equations of motion. 
	Simulations of the HMMA/graphene system were performed at constant temperature and constant pressure.
	Nos\'{e}-Hoover thermostat with $0.5$ ps relaxation time and Parrinello-Rahman barostat (with anisotropic pressure coupling) with $5$ ps relaxation time were used for controlling temperature and pressure.
	Depending on the temperature, we performed production runes for $100$ ns up to more than $4\ \mu$s.
	The nonbonded interactions were truncated at $1$ nm.  The particle-mesh Ewald method was employed for the calculation of the long-range electrostatic interactions. The tail of the 
	Van der Waals interactions was taken into account using either the analytical tail correction for energy and pressure or the particle-mesh Ewald method.  
	Excluding the simulations with adjusted liquid/surface attraction strength (discussed below), the Lorentz-Berthelot mixing rule was used for the calculation of the interactions between the graphene surface and the liquid atoms. 
	The employed force field can be found elsewhere~\cite{foroozani2021gradient}.

	The carbon atoms of graphene have weak attractive van der Waals interactions (modeled with the Lennard-Jones pair potential) with the atoms of HMMA. In the employed force field, the depth of the individual Lennard-Jones interactions between the carbon atoms of graphene and sp$^3$ carbon atoms of HMMA is approximately $0.35$ kJ/mol, smaller than the interaction energy between two sp$^3$ carbon atoms of HMMA (around $0.46$ kJ/mol).
	However, each atom of HMMA interacts with several carbon atoms of graphene, which have a packed honeycomb structure with a bond length of around $1.42$ \AA. 
	\autoref{Fig:model1}b shows a relaxed configuration of a single HMMA molecule on the graphene surface (the energetic cost of the separation of this configuration from the surface is around $\Delta E = 47$ kJ/mol). 
	\autoref{Fig:model1}c shows the 
	interfacial mass density profiles of the confined liquid at different temperatures.  
	The density profiles are calculated as a function of axial distance (the distance on the $z$ axis) from the center of mass of the confining surface. 
	Large local density oscillations, coming from the layered structure of the liquid, are observed in the vicinity of the surface. 
	With decreasing temperature, the layered structure propagates to longer distances.
	The temperature range shown in \autoref{Fig:model1}c ($350$ K - $180$ K) is the range in which the dynamics of the confined liquid is investigated. The \tg\ of the model liquid is estimated to be around $143$ K (see section S1 for the analysis of the structure and dynamics of the bulk liquid), and therefore the above temperature range approximately corresponds to $2.45$\tg\ down to $1.26$\tg. 
	The bulk liquid is studied in the range of $350$ K - $160$ K.

	\begin{figure}[!htb]
		\centering
		\begin{subfigure}{0.45\textwidth} 
			\includegraphics[width=\textwidth]{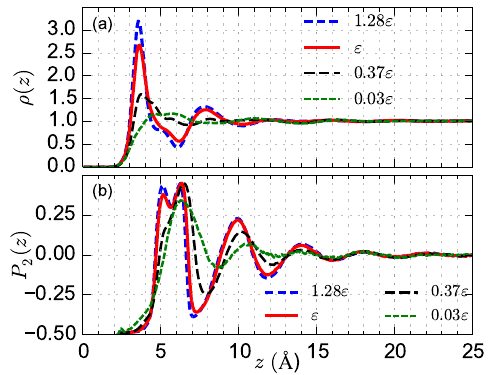}
		\end{subfigure}
		
		\caption{(a) Interfacial mass density profiles and (b) the profiles of the orientational order parameter, $P_2(z)$,  for the systems with different surface/liquid attraction strengths, adjusted through changing the depth of the individual Lennard-Jones interactions between the carbon atoms of graphene and the atoms HMMA. See the main text for the details. }
		
		\label{Fig:dens-eps} 
	\end{figure}

	To investigate the effect of the surface/liquid attraction 
	strength, we also performed three simulation runs, at $T = 200$ K, with 
	adjusted HMMA/graphene interactions (different from the reference system simulated using the original force field).
	In these simulations, the depths of the individual Lennard-Jones interactions, $\varepsilon$, between the carbon atoms of graphene and the atoms of HMMA were altered relative to the interactions in the reference system via the multiplication of the original $\varepsilon$ values by a constant factor. 
	The constant factor is $1.28$, $0.37$, or $0.03$. 
	When discussing the effect of the surface/liquid attraction strength, these systems with adjusted interactions are referred to with the labels $1.28 \varepsilon$, $0.37 \varepsilon$, and $0.03 \varepsilon$ and the reference system with the label $\varepsilon$.
	\autoref{Fig:dens-eps}a shows the local density profile of the  systems with the adjusted interactions. 
	With increasing the strength of the 
	attraction (higher $\varepsilon$ values), the
	layered structure of the liquid becomes more pronounced.
	In the case of the lowest attraction strength, the fluctuations of the local density are very small, resembling the behavior 
	in the vicinity of a wall with purely repulsive interactions with the liquid.
	
	Providing more details about the interfacial structure, 
	we present an analysis of the orientation of the HMMA molecules as a function of distance from the confining surface, through the measurement of the following order parameter: $P_2(z) = 3/2 \langle \cos^2 \theta(z) \rangle - 1/2$. Here $\theta$ shows the angle between a vector defined on the HMMA molecule (the vector between the $\alpha$ and ester carbons of HMMA as shown in \autoref{Fig:model1}b) and the confining surface. $P_2$ ranges between $-0.5$ (in the case of parallel orientation of the defined vectors relative to the surface) and $1$ (perpendicular orientation relative to the surface) and a value of $0$ shows random orientation of the defined vectors.   
	\autoref{Fig:dens-eps}b shows $P_2(z)$ for the systems with different surface/liquid attraction strengths ($T = 200$ K).
	Like density profiles, $P_2(z)$  has oscillatory behavior, originating from the interfacial layered structure of the liquid. 
	The maxima of the density profile almost correspond to the minima of $P_2(z)$ and vice versa. Hence, at the distances corresponding to the maxima of the density profile, the HMMA molecules prefer parallel orientations relative to the confining surface, and at the minima of the density profile, the molecules adopt partial perpendicular orientations. 
	In all systems, except the one with the lowest attraction strength ($0.03 \varepsilon$), at distances smaller than the first peak of the density profile the HMMA  molecules adopt almost perfect parallel orientation relative to the surface. 
	In the system with the lowest attraction, the first peak of the density profile is very weak and rather wide. In this case, the molecules that are very close to the surface have parallel orientation, however, the number of such molecules is significantly smaller than the system with higher attraction strength.

	\begin{figure}[!htb]
		\centering
		\begin{subfigure}{0.45\textwidth} 
			\includegraphics[width=\textwidth]{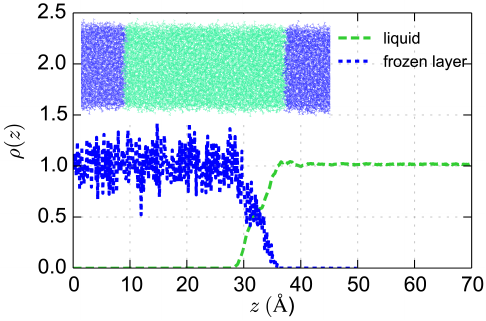}
		\end{subfigure}
		
		\caption{Mass density profiles for the case where the confining wall is a frozen amorphous layer of the bulk liquid ($T = 200$ K). The inset shows the studied geometry. }
		
		\label{Fig:dens-frozen} 
	\end{figure}

	As shown in \autoref{Fig:dens-eps}, in the vicinity of the graphene layer the structure of the liquid is different from the bulk structure. 
	Specifically, near graphene, the liquid forms a layered structure, and the liquid molecules adopt parallel to surface orientation. 
	To investigate the dynamics of the confined liquid in a situation in which the interfacial structure is similar to the bulk structure, we also 
	simulated HMMA liquid confined between frozen amorphous walls of the same liquid. 
	To prepare the frozen wall, an equilibrated configuration of the bulk liquid (equilibrated at the target temperature of the simulation) was taken and the centers of mass of all molecules of that configuration were wrapped inside its simulation box (no molecule was broken across the boundaries).
	The equations of motion were not solved for the atoms of the frozen wall. The simulations with the frozen configuration were performed at constant temperature ($200$ K and $350$ K) and constant volume.  
	The thickness of the frozen layer is around $6.4$ nm. 
	At $200$ K, the dimensions of the simulation box (containing the frozen wall and the liquid) are around $6.4 \times 6.4 \times 19.3$ nm$^3$.  
	\autoref{Fig:dens-frozen} shows the density profiles of the frozen wall and the confined liquid as functions of axial distance from the center-of-mass of the wall. 
	The frozen wall has a rough (nonplanar) surface 
	reflected in the unsharp decay of its density profile (it decays to zero within about $0.5$ nm). Furthermore, the density profile of the liquid does not show maxima and minima which is expected for a liquid in the vicinity of a configuration of itself.

	\section{Results and Discussion}
	
	We analyze the displacements of the particles of the liquid perpendicular to the confining surface (i.e., along the $z$ direction), by calculating self-intermediate scattering function ($F_\text{s} (q_z,t)$, wave vector is parallel to the $z$ axis), mean-squared displacement ($\langle \Delta z ^2(t) \rangle$), second-order non-Gaussian parameter ($\alpha_2^z$), and self part of the van-Hove function ($G_\text{s}(z,t)$), defined as:
	\begin{equation}
		\begin{aligned}	
			F_\text{s} (q_z,t) &= \langle \cos[q_z\Delta z(t)] \rangle\\
			\langle \Delta z ^2(t) \rangle &= \langle (z(t)-z(0))^2 \rangle \\
			\alpha_2^z(t) &= \frac{1}{3} \frac{\langle \Delta z ^4 \rangle}{\langle \Delta z ^2 \rangle^2}-1\\
			G_\text{s}(z,t) &=  \langle  \delta(z - [z_i(t) - z_i(0)]) \rangle
		\end{aligned}
		\label{Eq:dyn-quan}
	\end{equation}
	where $\langle \rangle$ shows averaging over appropriate atoms and time origins and $\delta(.)$ is the delta function.
	For the isotropic bulk liquid, the dynamical properties along all different directions are identical and the following relations can also be used:
	\begin{equation}
		\begin{aligned}	
			F_\text{s} (q,t) &= \langle \frac{\sin[q\Delta r(t)]}{q\Delta r(t)} \rangle 	  \\
			\langle \Delta r^2(t) \rangle &= 
			\langle ({\bf{r}}(t) - {\bf{r}}(0))^2 \rangle\\
			\alpha_2(t) &= \frac{3}{5} \frac{\langle \Delta r ^4 \rangle}{\langle \Delta r ^2 \rangle^2}-1\\
		\end{aligned}
		\label{Eq:SM-dynamics}
	\end{equation}
	where $\langle \Delta r^2(t) \rangle$ is the total mean-squared displacement of the particles which in the case of isotropic systems is equal to $3\langle \Delta z ^2(t) \rangle$.
	
	To track the displacements we look at the position of the $\alpha$-carbon atom of HMMA (see \autoref{Fig:model1}b).
	The dynamical quantities are calculated at different distances from the surface (layer-resolved). For the layer-resolved analysis, calculations are performed on the atoms present in a given layer at $t = 0$.
	
	\subsection{A Fickian liquid in the vicinity of a reflective surface}	
	
	\begin{figure}[!htb]
		\centering
		\begin{subfigure}{0.45\textwidth} 
			\includegraphics[width=\textwidth]{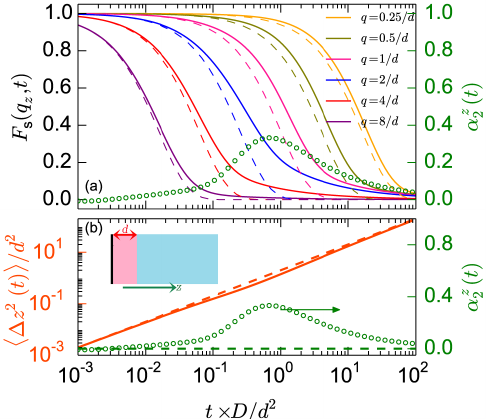}
		\end{subfigure}
		
		\caption{Solid curves and the curves with the circular marker: the analytically  calculated dynamical observables for a layer of a Fickian liquid 
			that at $t = 0$ has a thickness of $d$ and is located in contact with a reflective wall (see inset of panel b). The dashed curves display the observables for a bulk Fickian liquid.
		}
		\label{Fig:analytic} 
	\end{figure}
	
	Before presenting the simulation results, we examine the above-mentioned 
	dynamical 
	quantities for an idealized case of a liquid that exhibits Fickian diffusion in the vicinity of a reflective wall.
	In a bulk liquid, if the particles have Fickian diffusion then the distribution of their displacements is Gaussian:
	$G_\text{s}(z,t) = \exp(-z^2/4Dt)/\sqrt{4\pi Dt}$,
	$\alpha_2^z(t) = 0$, $\langle \Delta z ^2(t) \rangle = 2Dt$, and $F_\text{s} (q_z,t) = \exp(-q_z^2 D t)$, where $D$ is the diffusion coefficient of the particles of the liquid. 
	Now, consider a homogeneous layer of the liquid that at $t = 0$ has a thickness of $d$ and is located in contact with a reflective wall (see inset of \autoref{Fig:analytic}b).
	This geometry is relevant to the layer-resolved analysis of this work. 
	At the reflective wall $\partial c / \partial z = 0$, where $c$ is the concentration of the particles of the layer.
	The distribution of displacements for the particles close to a reflective wall can be calculated through the reflection of displacements at the wall~\cite{crank1979mathematics}. Through the reflection of displacements, the following relation for the $G_\text{s}(z,t)$ of the particles of the tagged layer (layer shown in \autoref{Fig:analytic}b) is obtained (see section S2 for more details):
	\begin{equation}
		\begin{aligned}
			G_\text{s}(z,t) &= \frac{1}{d\sqrt{4\pi Dt}}\int_{0}^{d} \big{[}\exp(-z^2/4Dt) +\\
			&\exp(-(z+2z_0)^2/4Dt)\big{]H(z+z_0)} \mathrm{d}z_0
			\label{Eq:reflective-wall-1}
		\end{aligned}
	\end{equation}
	where $H(.)$ is the unit (Heaviside) step function ($H(x) = 1$ for $x \geq 0$ and $H(x) = 0$ for $x < 0$).
	With $G_\text{s}(z,t)$ at hand, other dynamical observables can be calculated:
	\begin{equation}
		\begin{aligned}	
			F_\text{s} (q_z,t) &=  = \int_{-\infty}^{\infty} G_\text{s}(z,t) \cos[q_z z] \mathrm{d}z 
			\\
			\langle \Delta z ^2(t) \rangle &=  \int_{-\infty}^{\infty} G_\text{s}(z,t) z^2 \mathrm{d}z   \\
			\langle \Delta z ^4(t) \rangle &= \int_{-\infty}^{\infty} G_\text{s}(z,t) z^4 \mathrm{d}z   \\
		\end{aligned}
		\label{Eq:reflective-wall-2}
	\end{equation}
	
	We calculated the above integrals and that of \autoref{Eq:reflective-wall-1} numerically. 
	The calculated 
	dynamical quantities are shown in  \autoref{Fig:analytic}.
	The dynamical observables for the particles of the tagged layer deviate from the bulk ones. This is expected because  a particle 
	that	moves in the negative $z$ direction near the wall has to come back to its original location.  
	The $\alpha_2^z(t)$ of the layer shows a peak.
	The $F_\text{s} (q_z,t)$ of the layer decays slower than the corresponding bulk curve and exhibits a tail; the deviation is  pronounced for $q_z $ values around $0.5/d$ to $4/d$.       
	The $\langle \Delta z ^2(t) \rangle$  of the layer has an angle around the peak time of $\alpha_2^z(t)$. Around this time, $\langle \Delta z ^2(t) \rangle$ first shows a mild sub-diffusive and then a mild super-diffusive behavior.  	These calculations are useful for understanding the mere geometric effect of confinement on the dynamical observables of the liquid, as discussed below. 
	They also explain the non-monotonic behavior of $F_\text{s} (q_z,t)$ in the case of layer-resolved analysis with very thin layers, as discussed in section S2.  
	
	\subsection{Simulation results}
	
	\subsubsection{HMMA/graphene system}
	
\begin{figure}[!htb]
		\centering
		\begin{subfigure}{0.45\textwidth} 
			\includegraphics[width=\textwidth]{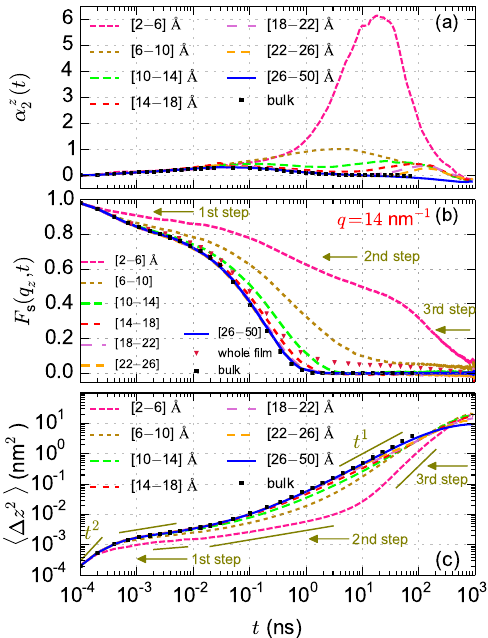}
		\end{subfigure}
		\caption{The layer-resolved $\alpha_2^z(t)$, $F_\text{s} (q_z,t)$,  and $\langle \Delta z ^2(t) \rangle$ curves for characterizing the displacements of the particles perpendicular to the surface ($T = 200$ K). 
			In panel (c) the tangent lines show the changes in the slope of the curve. 
		}
		\label{Fig:layer-resolved} 
	\end{figure}

	In simulations, for the layer-resolved analysis of the dynamics, the confined film is divided into seven layers.
	Close to the surface, we define six layers of thickness of $4$ \AA; however, for increasing visibility, we consider the center of the film ($[26$--$50]$ \AA\ from the surface) as a single layer (see the legends of \autoref{Fig:layer-resolved}).
	The selected $4$ \AA\ thickness corresponds to the 
	length scale of the local density oscillations close to the graphene sheet (see \autoref{Fig:model1}c). 
	Particularly, the first layer corresponds to the width of the first peak of the density profile (distance up to the first minimum of the density profile).

	\autoref{Fig:layer-resolved}a shows the
	layer-resolved $\alpha_2^z(t)$ curves at $T = 200$ K, roughly corresponding to $T = 1.4$\tg.
	The $\alpha_2^z(t)$ of the bulk liquid and of the central layers of the confined liquid exhibit a peak at short times (around $50$ ps at $200$ K). This peak is related to the dynamical heterogeneity originated from the cage effect (see also section S1 and Figure S2).
	The  $\alpha_2^z(t)$ of the first layer starts to increase almost in parallel to the bulk $\alpha_2^z(t)$.
	However, after the peak time of the $\alpha_2^z(t)$ of the bulk liquid, the $\alpha_2^z(t)$ of the first layer continues to increase and shows a  strong peak at later times.
	This peak is much stronger than the peak of the $\alpha_2^z(t)$ close to a reflective wall (\autoref{Fig:analytic}); hence, it is not a mere geometric effect of the confining surface.   
	This strong peak is related to the trapping of the liquid particles in the vicinity of the graphene surface.
	At short times most of the particles, which were trapped near the surface at $t = 0$ stay in the vicinity of the surface; at intermediate times, some particles escape from the surface 
	but some of them are still trapped near the surface; this leads to heterogeneity of displacements observed through the peak of $\alpha_2^z(t)$.
	The $\alpha_2^z(t)$ of the second layer also has
	a considerable peak, however, it is much lower than that of
	the first layer. Further, at long times ($20$ -- $200$ ns in \autoref{Fig:layer-resolved}a) small peaks are observed in the $\alpha_2^z(t)$ of the third to sixth layers. 
	The particles belonging to these layers are far from the surface at $t = 0$;
	after some time, some of those  come close to the  surface and feel its effects. 
	
	\autoref{Fig:layer-resolved}b shows the layer-resolved $F_\text{s} (q_z,t)$ functions at $T = 200$ K at $q_z = 14$ nm$^{-1}$, which is close to the peak position of the structure factor of the $\alpha$-carbon atoms of HMMA (Figure S1).
	The $F_\text{s} (q_z,t)$ of the bulk liquid and of the central layers shows its typical behavior for a glass-forming liquid at a rather high temperature \cite{binder2011glassy}; it shows a two-step decay, however, the temperature is not low enough to see the full separation between the two steps.
	At $T = 200$ K,
	the surface modifies the  $F_\text{s} (q_z,t)$ of the four adjacent layers.
	With reducing temperature, the effect of the surface on the liquid dynamics propagates to longer distances. For example at $T = 280$ K, only the first and second layers  are affected (see section S3).

	The $F_\text{s} (q_z,t)$ of the first layer decays  slower than that of the bulk liquid. 
	Furthermore, it exhibits a three-step decay (\autoref{Fig:layer-resolved}b).
	Similar to the bulk liquid, the first step of the $F_\text{s} (q_z,t)$ of the first layer is related to
	the rattling of the particles in the cages formed by their neighbors (due to vibrational motion of the atoms in the molecule and restricted movement of the molecules in the cages formed by other molecules).
	The decay of the $F_\text{s} (q_z,t)$ of the first layer in its first step is smaller than the decay of the bulk relaxation function in the corresponding step. 
	After this microscopic regime,  
	the particles only have small-scale back-and-forth movements close to the surface. This is reflected in the partial decay of $F_\text{s} (q_z,t)$  in its second step.
	Considering the parallel-to-surface orientation of the molecules in the first layer (\autoref{Fig:dens-eps}), 
	in the second step of $F_\text{s} (q_z,t)$, the liquid molecules probably have seesaw-like movements on the graphene surface, however, because of the attraction of the surface they can not escape the surface. 
	At later times, when the molecules
	start to escape from the surface, an additional relaxation step (the third step)  emerges in the $F_\text{s} (q_z,t)$ curve.
	The onset of the third step corresponds to the position of the large peak of the $\alpha_2^z(t)$ of the first layer.    
	For the bulk liquid, the onset of the second step of $F_\text{s} (q,t)$ also coincides with the peak of the bulk $\alpha_2(t)$ (see Figure S2). \autoref{Fig:layer-resolved}b shows also $F_\text{s} (q_z,t)$  for the whole confined liquid. This quantity can be measured experimentally, e.g., through quasi-elastic neutron scattering~\cite{krutyeva2009neutron}. Note that $F_\text{s}^\text{whole} (q_z,t) = \sum \phi_i\ F_{\text{s}_{i}} (q_z,t)$, where $\phi_i$ and $F_{\text{s}_{i}} (q_z,t)$ are the fraction of the particles present in  layer $i$ and its scattering function.

\begin{figure}[!htb]
		\centering
		\begin{subfigure}{0.45\textwidth} 
			\includegraphics[width=\textwidth]{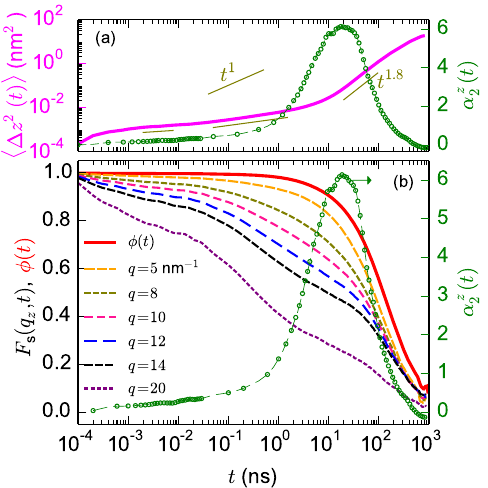}
		\end{subfigure}
		
		\caption{$\langle \Delta z ^2(t) \rangle$ and $F_\text{s} (q_z,t)$   of the first layer together with its $\alpha_2^z(t)$. $\phi(t)$ in panel (b) is the fraction of the trapped particles at $t = 0$ which are still trapped (near the same surface) at time $t$ ($T = 200$ K).
			\label{Fig:layer1}} 
	\end{figure}

	{\jcp \autoref{Fig:layer-resolved}c shows the layer-resolved $\langle \Delta z ^2(t) \rangle$ curves at $T = 200$ K. Because of confinement in a finite volume, at long times, the $\langle \Delta z ^2(t) \rangle$ curves of all layers become constant. 
		After the short-time ballistic regime ($\langle \Delta z ^2(t) \rangle \propto t^2$) the  $\langle \Delta z ^2(t) \rangle$ of the bulk liquid (and of the central layers) exhibits a sub-diffusive regime and then converges to normal diffusion ($\langle \Delta z ^2(t) \rangle \propto t^1$). 
		This sub-diffusive regime is related to the cage effect, however, at the studied temperature, $200$ K, the cages are not rigid enough to see a plateau~\cite{binder2011glassy}.
		After the ballistic regime, the $\langle \Delta z ^2(t) \rangle$ of the first layer shows two sub-diffusive regimes (with different slopes). 
		Similar to the bulk liquid, the first sub-diffusive regime comes from the cage effect. The second regime comes from the trapping effect of the surface; this regime corresponds to the second step of the decay of $F_\text{s} (q_z,t)$.
		After these sub-diffusive regimes,  $\langle \Delta z ^2(t) \rangle$ increases rapidly and shows a pronounced super-diffusive behavior. 
		The onset of the super-diffusive regime almost concurs with the position of the peak of $\alpha_2^z(t)$ and the onset of the third step of $F_\text{s} (q_z,t)$ (see \autoref{Fig:layer1}). This regime is related to the process of escaping from the surface.
		As shown in \autoref{Fig:analytic}b, a super-diffusive behavior is also observed  
		close to a reflective wall. However, the effect is much stronger for the model system.
		To explain the super-diffusive behavior, 
		we should note that for the calculation of $\langle \Delta z ^2(t) \rangle$
		of the first layer, particles present in the layer at $t = 0$ are considered, and displacements are calculated relative to the initial position of the particles in the layer.  
		After escaping from the surface, 
		the particles do not have a super-diffusive motion (look at the $\langle \Delta z ^2(t) \rangle$ of the central layers), however, $\langle \Delta z ^2(t) \rangle$
		calculated relative to the initial positions of the particles in the vicinity of the surface, increases rapidly, and exhibits an \emph{apparent} super-diffusive behavior. Note that $\langle \Delta z ^2(t) \rangle$ is dominated by the fast particles.
		To support the above explanation about the super-diffusive regime, we have also calculated
		$\langle \Delta z ^2(t) \rangle = \langle (z(t_\text{w} + t) - z(t_\text{w}))^2 \rangle$ for the particles of the first layer (particles present in the first layer at time $t = 0$)
		after different waiting times, $t_\text{w}$ (the results are shown in Figure S7). With increasing $t_\text{w}$, the super-diffusive regime gradually disappears, which shows that
		the particles do no have super-diffusive behavior after escaping from the surface (for a more detailed discussion see section S4).}	
	
	{\jcp For the above discussion, the layer-resolved $F_\text{s} (q_z,t)$ functions was shown at  $q = 14$ nm$^{-1}$. Considering the large heterogeneity of the displacements of the particles of the first layer, it is informative to investigate the $q$-dependence of the $F_\text{s} (q_z,t)$ of this layer}.
	\autoref{Fig:layer1}b shows the $q$-dependence of the $F_\text{s} (q_z,t)$ of the first layer at $200$ K. 
	In a rather wide range of $q$, 
	$F_\text{s} (q_z,t)$ vanishes at a $q$-independent  time, after the peak time of  $\alpha_2^z(t)$.
	Such a $q$-dependence resembles the behavior of the single-chain scattering function of entangled polymer melts, which according to the tube model is governed by two processes: local chain reptation
	inside the tube and the escape of the chain from its confining tube\cite{deGennes1981coherent}. 
	To understand this $q$-dependence, we follow the argument made in \cite{deGennes1981coherent}.
	A particle contributes to  $F_\text{s} (q_z,t)$, if its displacement  $\Delta z(t) \lesssim 2\pi q_z^{-1}$ (i.e., it is slow).
	The particles that escape from the surface become mobile and, if $q_z$ is not very small,  translate by distances larger than $2\pi q_z^{-1}$ while some other particles are still trapped near the surface.
	These mobile particles do not contribute to the scattering function of the first layer and  $F_\text{s} (q_z,t)$
	is dominated by the particles which are still trapped near the surface.
	Based on this argument, for a range of $q_z$, the long-time behavior of $F_\text{s} (q_z,t)$ of the first layer is controlled 
	by $\phi(t)$ which is the fraction of the trapped particles (scattering centers) at $t = 0$ that are still trapped there (near the same surface) at time $t$.
	We assume that a scattering center is trapped near the surface if it is located in the first layer. 
	Such particles 
	are also described as adsorbed or attached~\cite{smith2005polymer,behbahani2020conformations}; also since $\phi(t)$ can be calculated like a correlation function~\cite{yelash2010slow,yelash2012three,solar2017relaxation,behbahani2020conformations}, it is usually called desorption correlation function.
	It can be calculated using:
	\begin{equation}
		\phi(t) = \langle S(t) S(0) \rangle = \langle S(t) \rangle  \end{equation}
	where $S(t) \in \{0, 1\}$ is assigned to each particle ($\alpha$-carbon atoms here) as follows: $S(t) = 1$ if at time $t$, the particle is adsorbed on the same surface that it was adsorbed at $t = 0$; otherwise $S(t) = 0$. $\langle \rangle$ shows averaging over time origins and the particles that are adsorbed at $t = 0$ (i.e., $S(0) = 1$). 
	Because of the confinement in a finite volume, $\phi(t)$ does not decay to zero at long times, instead, converges to the fraction of all particles of the liquid which are trapped near one side of the surface (around $0.045$ in the current geometry). $\phi(t)$ can be easily scaled to 
	decay to zero at long times\cite{yelash2010slow,yelash2012three,solar2017relaxation}, however, we do not scale it, because the finiteness of the confinement volume also affects $F_\text{s} (q_z,t)$ and $\Delta z ^2(t) \rangle$ (see below).
	$\phi(t)$  at $200$ K is shown in \autoref{Fig:layer1}b.
	The final decay time of the $F_\text{s} (q_z,t)$ of the first layer over a wide $q$ range is rather close to the decay time of $\phi(t)$, which is consistent with the above discussion.
	The development of the large peak of $\alpha_2^z(t)$ is also in line with the decay of $\phi(t)$. This is consistent with the discussion of \autoref{Fig:layer-resolved}a regarding the relation of this peak to the process of escaping from the surface.

	\begin{figure}[!htb]
		\centering
		\begin{subfigure}{0.45\textwidth} 
			\includegraphics[width=\textwidth]{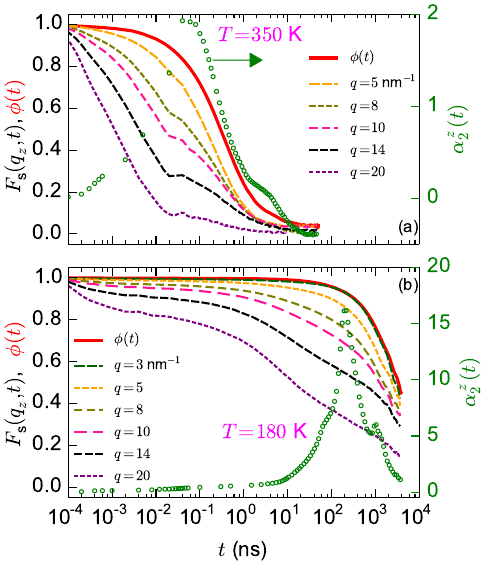}
		\end{subfigure}
		
		\caption{The $q$-dependence of the  $F_\text{s}(q,t)$ of the first layer  at (a) $T = 350$ K and (b) $T = 180$ K. In each panel, $\phi(t)$ and
			$\alpha_2^z(t)$ of the first layer are also presented. 
		}
		\label{Fig:Fs-qs} 
	\end{figure}

	\autoref{Fig:Fs-qs} shows the $q$-dependence of the $F_\text{s} (q_z,t)$ of the first layer at $350$ K and $180$ K. With decreasing temperature, the long-time convergence of the $F_\text{s} (q_z,t)$ curves (over a range $q$), and the long-time convergence of $F_\text{s} (q_z,t)$ and $\phi(t)$, become more pronounced. 
	This stems from the higher heterogeneity of particle displacements at lower temperatures. 
	With increasing dynamic heterogeneity, the distribution of particle displacements becomes more consistent with the categorization of the particles as trapped (near the surface) and mobile, which is the basis of the argument for the long-time convergence of  $F_\text{s} (q_z,t)$. 
	The temperature dependencies of dynamic heterogeneity and distribution of particle displacements are discussed below.

	\begin{figure}[!htb]
		\centering
		\begin{subfigure}{0.45\textwidth} 
			\includegraphics[width=\textwidth]{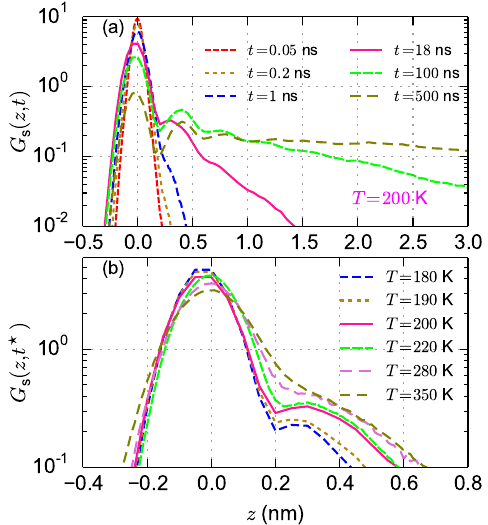}
		\end{subfigure}
		
		\caption{$G_\text{s}(z,t)$, for the particles that are present in the first layer at time origin. 
			(a) $G_\text{s}(z,t)$ at $T = 200$ K at different times.
			(b) $G_\text{s}(z,t^\star)$ at different temperatures. 
			$t^\star$ is the peak time of $\alpha_2^z(t)$.  
		}
		\label{Fig:Gs} 
	\end{figure}

	{\jcp Given the non-Gaussian distribution of particle displacements (deduced from the large peak of the non-Gaussian parameter for the particles of the first layer), it is instructive to look at the self-part of the Van Hove function, $G_\text{s}(z,t)$, which explicitly shows the distribution of displacements along the $z$ direction.}	
	\autoref{Fig:Gs} shows $G_\text{s}(z,t)$ for the particles that are present in the first layer at $t = 0$. Note that, for the averaging over both sides of the confining surface,  the distribution of displacements for the left-hand side particles has been reflected over $z = 0$ and then the average distribution has been calculated. 
	\autoref{Fig:Gs}a shows the $G_\text{s}(z,t)$ at different times at $T = 200$ K.
	At short times, the particles have small movements within the layer (probably due to the seesaw-like motion of molecules near the graphene surface) and $G_\text{s}(z,t)$ has an almost symmetric shape. Gradually some particles leave the layer and diffuse toward the center of the film. At later times a secondary peak is seen in $G_\text{s}(z,t)$. This secondary peak originates from the layered structure of the liquid close to the surface.

	\begin{figure}[!htb]
		\centering
		\begin{subfigure}{0.45\textwidth} 
			\includegraphics[width=\textwidth]{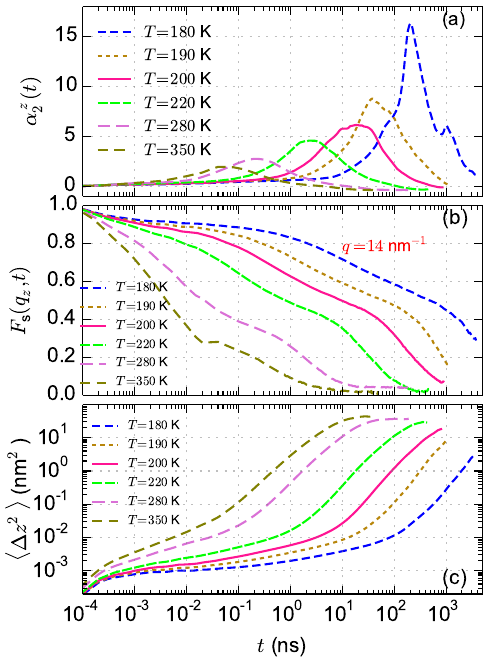}
		\end{subfigure}
		
		\caption{Temperature dependence of different dynamical quantities for the particles present in the first layer at time origin.  }
		\label{Fig:layer1-Ts} 
	\end{figure}

	\autoref{Fig:layer1-Ts} shows the temperature dependence of the dynamical observables of the first layer.
	The presented temperature range approximately corresponds to $2.45$\tg\ to $1.26$\tg. 
	The $\alpha_2^z(t)$ curves (panel a) 
	show strong peaks because of the surface effect on the displacement of particles.
	With decreasing temperature, the heterogeneity of the particle displacements significantly increases, reflected in the higher
	peaks of the $\alpha_2^z(t)$.
	\autoref{Fig:Gs}b presents $G_\text{s}(z,t)$ curves at the peak time of $\alpha_2^z(t)$ ($t = t^{\star}$). 
	At all temperatures $G_\text{s}(z,t^{\star})$ exhibits a shoulder which is related to the displacement of the  particles that have escaped from the surface. With reducing temperature
	the height of the main peak of $G_\text{s}(z,t^{\star})$ increases. This large peak is related to the slow particles that are still trapped near the surface. Furthermore, the 
	shoulder of $G_\text{s}(z,t^{\star})$  becomes more pronounced and appears as a small secondary peak.
	In Figure S8 we also show 
	$\alpha_2^z(t)$ vs $\Delta z^2(t)$ for the particle of the first layer.  
	In the studied temperature range the peak of $\alpha_2^z(t)$ appears around $2.5$--$4.3$ \AA$^2$.
	Upon reducing temperature, the position of the peak shifts to smaller values of $\Delta z^2(t)$.
	These length scales correspond to the small-scale movements of the scattering centers in the interfacial layer.

	\autoref{Fig:layer1-Ts}b shows the $F_\text{s} (q_z,t)$ of the first layer at different temperatures.
	At high temperatures, the cages are not rigid and their effect is not clearly seen (i.e., at short times $F_\text{s} (q_z,t)$ does not have two distinct steps).
	As for bulk glass-forming liquids, upon reducing temperature, the time scales of the first and second steps of $F_\text{s} (q,t)$ become more separated. This is reflected in the appearance of an extended plateau between the first and second steps.
	With reducing temperature, such a sign of time scale separation between the second and the third steps of $F_\text{s} (q_z,t)$ of the first layer is not observed; on the contrary, the separation between the second and third steps becomes less clear and their time scales seem to come closer to each other.
	Furthermore, the contribution of the third step in the decay of $F_\text{s} (q_z,t)$ seems to increase; that is, the onset of the third step appears at larger $F_\text{s}$ values.

\begin{figure}[!htb]
		\centering
		\begin{subfigure}{0.45\textwidth} 
			\includegraphics[width=\textwidth]{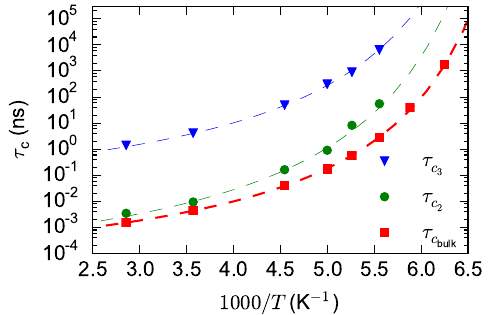}
		\end{subfigure}
		
		\caption{The estimated time scales of the second and third steps of the $F_\text{s} (q_z,t)$ of the first layer.}
		\label{Fig:VFT} 
	\end{figure}

	The  $F_\text{s} (q_z,t)$ curves have also been analyzed by fitting their second and third steps (i.e., excluding the short time fast motions) with the sum of two stretched-exponential functions: $f(t) = A_2 \exp[ -({t}/\tau_2)^{\beta_2}] + A_3 \exp[ -({t}/\tau_3)^{\beta_3}]\ (A_2 + A_3 < 1) $. 
	The time scale associated with each step has been calculated using $\tau_{c_i} = \tau_i/\beta_i \Gamma(1/\beta_i)$, where $\Gamma(.)$ is the gamma function.   
	The values of $\tau_{c_i}$  at different temperatures are provided in \autoref{Fig:VFT} and they are in favor of the gradual merging of the second and third steps with decreasing temperature. 
	To better see the trend, we also fitted Vogel-Fulcher-Tammann (VFT) functions ($\tau_c = \tau_\infty \exp(B/(T - T_0))$) on the time scales of the second and third steps. The fitted curves are shown as thin dashed curves in \autoref{Fig:VFT}.
	The parameters of the fitted VFT curves are:  
	$T_0 = 129.0$ K, $B = 500$ K, and $\tau_\infty = 0.00016$ ns for bulk.
	$T_0 = 130.0$ K, $B = 620$ K, and $\tau_\infty = 0.00016$ ns for the second step and $T_0 = 130.0$ K, $B = 555$ K, and $\tau_\infty = 0.1$ ns for the third one.
	It should be noted that the accuracy of the fitted VFT functions on the time scales of the first layer is lower than the VFT function fitted on the bulk relaxation times. 
	Because the relaxation times of the first layer are outside of the simulation window at $170$ K and $160$ K; even at $180$ K full decay of the relaxation function can not be seen, despite more than $4\ \mu$s run.

	It is worth noting that after the third step, $F_s(q_z,t)$ curves do not decay completely and have residues, appearing as long-time plateaus (\autoref{Fig:layer1-Ts}b). The appearance of these plateaus almost concurs with the convergence of $\langle \Delta z ^2(t) \rangle$ to a constant value 
	and it is related to the confinement of the liquid in a finite volume.

	\autoref{Fig:layer1-Ts}c shows the $\langle \Delta z ^2(t) \rangle$ of the first layer at different temperatures.
	At low temperatures, 
	two sub-diffusive regimes can be detected (which are discussed above).
	At high temperatures, the cages are not rigid enough and their restricting effects are not seen; hence only one sub-diffusive regime is observed.
	Furthermore, at lower temperatures, transition to the super-diffusive regime occurs at smaller $\langle \Delta z ^2(t) \rangle$ values.
	This is consistent with the smaller contribution of the second step in the decay of $F_\text{s} (q_z,t)$ at lower temperatures.
	With decreasing temperature, the slope of the second sub-diffusive regime gradually reduces and the slope of the apparent super-diffusive regime seems to gradually increase. 
	
	Generally, with increasing temperature, the behaviors of the dynamical quantities become closer to the behaviors observed in the vicinity of a reflective wall (\autoref{Fig:analytic}): At higher temperatures, the peak height of $\alpha_2^z(t)$ is smaller, the apparent super-diffusive behavior of $\langle \Delta z ^2(t) \rangle$ is weaker, and the long-time convergence  of    
	$F_\text{s} (q_z,t)$ curves (over a range of $q$) is less pronounced. 
	However, even at the highest studied temperature, the shape of the measured $F_\text{s} (q_z,t)$ curve is different from the idealized scattering function near a reflective wall.  
	This difference probably originates from the layered structure of the liquid and the specific orientation of the molecules near the graphene surface.

	\subsubsection{The effect of the surface/liquid attraction strength}
	
	\begin{figure}[!htb]
		\centering
		\begin{subfigure}{0.45\textwidth} 
			\includegraphics[width=\textwidth]{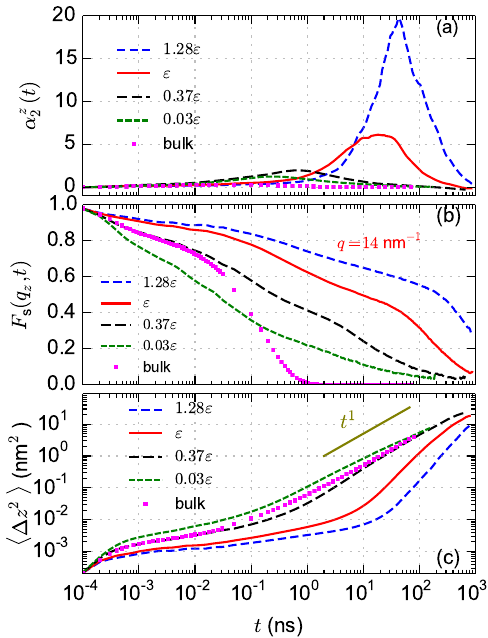}
		\end{subfigure}
		
		\caption{The effect of the surface/liquid attraction strength on the dynamical observables of the particles present in the first layer at time origin 
			($T = 200$ K). 
			The corresponding interfacial density profiles are shown in \autoref{Fig:dens-eps}. }
		\label{Fig:1layer-eps} 
	\end{figure}
	
	{\jcp In this section, we present the dynamical observables of the first interfacial layer in systems with different surface/liquid attraction strengths. 
		The effect of the surface/liquid attraction strength on the dynamics of confined liquids has been studied before in the literature~\cite{zhang2017effects,zhang2018we}.
		In these works, a slowing down of the relaxation dynamics of the neighboring liquid was reported with increasing attraction of the surface.
		In this section, however, we focus on the analysis of the signatures of the process of escaping from the surface in systems with different attraction strengths. 
		\autoref{Fig:1layer-eps} shows the effect of the attraction strength (of graphene/HMMA) on the dynamics of the first layer ($T = 200$ K).}
	The corresponding density profiles are provided in \autoref{Fig:dens-eps}.  
	The reference system, which is extensively studied above, is denoted by $\varepsilon$ in this figure.
	As mentioned in \autoref{sec:model}, in the systems that are denoted by $k\varepsilon$ ($k$ is a numerical factor), the depths of the individual 
	Lennard-Jones interactions between the carbon atoms of graphene and the atoms of HMMA are $k$ times the depths of the interactions in the reference system.
	The dynamical quantities of the bulk liquid at $200$ K are also shown in \autoref{Fig:1layer-eps}.
	With increasing the attraction strength, the surface-induced dynamical heterogeneity increases, reflected in the larger peak of $\alpha_2^z(t)$.  
	Furthermore, the sign of time scale separation between the second and third steps of $F_\text{s} (q_z,t)$ is not observed, and the slope of the apparent super-diffusive
	regime of  $\langle \Delta z ^2(t) \rangle$ increases upon the rise of the attraction strength.
	These trends are qualitatively similar to those observed upon reducing temperature (see \autoref{Fig:layer1-Ts}).
	The behavior of $F_\text{s} (q_z,t)$ at short and intermediate time scales (before the appearance of the third step) is also worth attention. At these time scales, with decreasing the attraction strength of the surface, the decay of $F_\text{s} (q_z,t)$ of the first layer becomes faster; particularly, in the system with the lowest attraction strength ($0.03 \varepsilon$), the decay is faster than the bulk behavior (at short and intermediate times).    
	In this system, the increase of $\langle \Delta z ^2(t) \rangle$ of the first layer is also faster than the bulk behavior.  
	It is worth noting that the dynamics of the liquid parallel to the surface depends also on the attraction strength and in $0.03 \varepsilon$ and $0.37 \varepsilon$ systems is faster than the bulk dynamics (data are not shown).

	\begin{figure}[!htb]
		\centering
		\begin{subfigure}{0.45\textwidth} 
			\includegraphics[width=\textwidth]{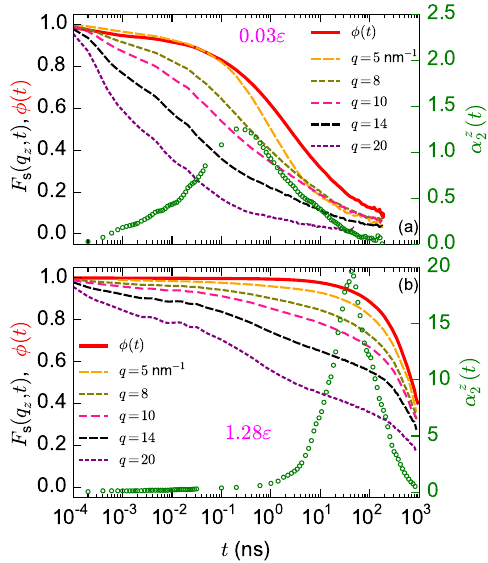}
		\end{subfigure}
		
		\caption{The $q$-dependence of the  $F_\text{s}(q,t)$ of the first layer for systems with different surface/liquid interaction strengths ($T = 200$ K).  
			In each panel, $\phi(t)$ and
			$\alpha_2^z(t)$ of the first layer are also presented. 
		}
		\label{Fig:qs-eps} 
	\end{figure}

	\autoref{Fig:qs-eps} shows the $q$ dependence of the $F_\text{s}(q,t)$ of the first layer in $0.03\ \varepsilon$ and $1.28\ \varepsilon$ system (the lowest and the highest attraction strengths studied here).
	The behavior of $F_\text{s}(q,t)$  in the $0.03\ \varepsilon$ system resembles the behavior observed in the vicinity of a reflective wall. In contrast, in the $1.28\ \varepsilon$ system, a long-time convergence of the $\phi(t)$ function and the $F_\text{s}(q,t)$ curves is observed over a wide range of $q$. The convergence originates from the strong heterogeneity of the particle displacements (visible in the large peak of $\alpha_2(t)$) and the dominance of slow particles over the decay of $F_\text{s}(q,t)$ (see the discussion of \autoref{Fig:Fs-qs}).        
	
	\subsubsection{A frozen configuration of the liquid as the confining wall}

	\begin{figure}[!htb]
		\centering
		\begin{subfigure}{0.45\textwidth} 
			\includegraphics[width=\textwidth]{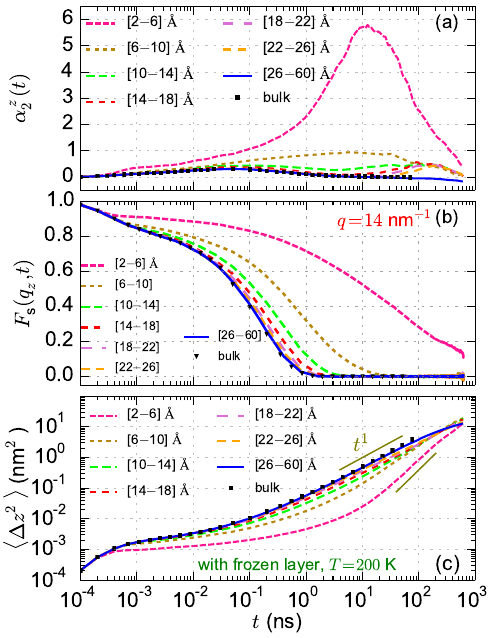}
		\end{subfigure}
		
		\caption{The layer-resolved $\alpha_2^z(t)$, $F_\text{s} (q_z,t)$,  and $\langle \Delta z ^2(t) \rangle$ curves for the liquid confined between frozen walls (frozen configurations of the bulk liquid) at $T = 200$ K.}
		\label{Fig:frozen-200} 
	\end{figure}

	As shown in \autoref{Fig:model1} and \autoref{Fig:dens-eps},  in the vicinity of the graphene surface, the structure of the liquid is different from the bulk structure. Specifically, near the graphene surface, the liquid forms a layered structure, and HMMA molecules adopt parallel-to-surface orientation. 
	To study the relaxation dynamics of the interfacial layer in a situation in which the interfacial structure is similar to the bulk structure, we study HMMA confined between frozen amorphous configurations of the bulk liquid. 
	In this situation, the interfacial structure is similar to the bulk structure (see \autoref{Fig:dens-frozen}), and also the surface/liquid interaction is similar to the liquid/liquid interaction.
	Even though the liquid does not form a layered structure near the frozen wall, we define different layers near the frozen wall (similar to those used for the layer-resolved analysis of the dynamics near the graphene surface) and calculate dynamical quantities at different layers separately.  
	Considering the roughness of the frozen wall, a layer is assigned to a scattering center ($\alpha$-carbon atoms) based on its minimum distance to the atoms of the wall.
	\autoref{Fig:frozen-200} shows the layer-resolved dynamical quantities of the liquid confined between frozen configurations, at $T = 200$ K.
	Despite the absence of the interfacial layered structure,
	in the vicinity of the frozen wall some features of the dynamical quantities are similar to those near the graphene surface (\autoref{Fig:layer-resolved}). Specifically, 
	the $\alpha_2^z(t)$ of the first layer exhibits a  large peak, and around the peak time of $\alpha_2^z(t)$,    
	$\langle \Delta z ^2(t) \rangle$ increases rapidly and shows an apparent super-diffusive behavior.
	The $F_\text{s}(q,t)$ of the first layer is stretched, however, near the frozen wall, it does not show a clear sign of three-step decay. Here, because of the absence of the layered structure and surface-induced alignment of the molecules, the small-scale movements of molecules near the wall and the process of escaping from the wall are not well-separated from each other and $F_\text{s}(q,t)$ does not show a clear three-step decay.

	\begin{figure}[!htb]
		\centering
		\begin{subfigure}{0.45\textwidth} 
			\includegraphics[width=\textwidth]{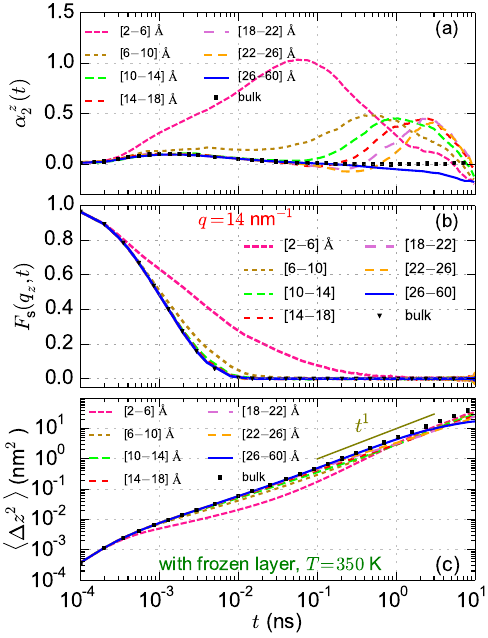}
		\end{subfigure}
		
		\caption{The layer-resolved $\alpha_2^z(t)$, $F_\text{s} (q_z,t)$,  and $\langle \Delta z ^2(t) \rangle$ curves for the liquid confined between frozen walls (frozen configurations of the bulk liquid) at $T = 350$ K.}
		\label{Fig:frozen-350} 
	\end{figure}

	\autoref{Fig:frozen-350} shows the dynamical observables in the vicinity of the frozen wall at $T = 350$ K. Comparing \autoref{Fig:frozen-350} with \autoref{Fig:frozen-200} shows that similar to the case of confinement between graphene surfaces, upon increasing temperature the peak height of $\alpha_2^z(t)$ of the first layer decreases, and the apparent super-diffusive regime of $\langle \Delta z ^2(t) \rangle$ (of the first layer) becomes less pronounced.  
	At $350$ K, the $F_\text{s}(q,t)$ of the first layer has a long tail which resembles the behavior of $F_\text{s}(q,t)$ near a reflective wall (\autoref{Fig:analytic}). 
	$F_\text{s}(q,t)$ shows also a long tail near the graphene surface at $350$ K (see \autoref{Fig:layer1-Ts}); however, unlike the behavior of $F_\text{s}(q,t)$ near the frozen wall,  near the graphene surface 
	a clear separation between the 
	short-time decay of $F_\text{s}(q,t)$  and its long-time tail is observed.   
	As discussed above, this difference probably originates from the preferential alignment of the HMMA molecules in the vicinity of the graphene surface.

	The length scale over which the frozen wall affects the dynamics of the adjacent liquid is also worth attention. At $350$ K, the dynamics in the first and second regions (up to $1$ nm distance) is different from the bulk behavior. Upon decreasing temperature, the effect of the wall reaches further regions; at $200$ K, up to the six neighboring regions (up to $2.5$ nm distance) are affected. Such 
	behavior is also observed in the vicinity of the graphene surface (see Figure S6).
	A qualitatively similar trend has also been observed for a binary Lennard-Jones liquid in the vicinity of a wall that has been realized by a frozen configuration of that liquid~\cite{scheidler2004relaxation} (similar to the system studied here).
	This behavior has been explained by referring to the increase of the length scale of cooperative motions of the particles of liquid with decreasing temperature~\cite{scheidler2004relaxation}.

	\begin{figure}[!htb]
		\centering
		\begin{subfigure}{0.45\textwidth} 
			\includegraphics[width=\textwidth]{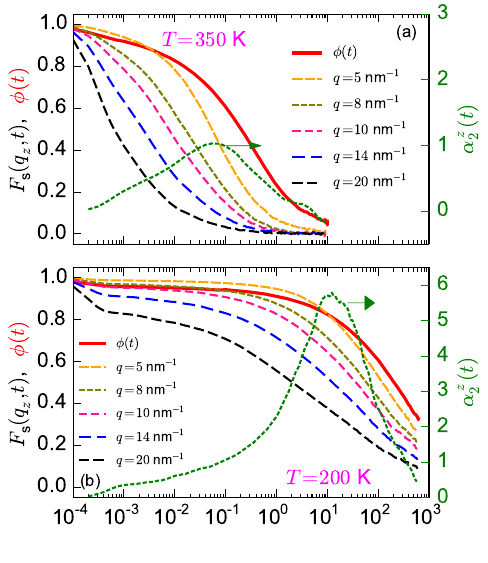}
		\end{subfigure}
		
		\caption{The $q$-dependence of the $F_\text{s}(q,t)$ of the first region in the vicinity of the frozen wall,  at (a) $T = 350$ K and (b) $T = 280$ K. $\alpha_2^z(t)$ of the first layer are also presented.}
		\label{Fig:qs-eps-frozen} 
	\end{figure}
	
	\autoref{Fig:qs-eps-frozen} shows the $q$ dependence of the $F_\text{s}(q,t)$ of the first interfacial region in the vicinity of the frozen wall. Panel (a) and panel (b) show the $q$-dependence of $F_\text{s}(q,t)$ at $350$ K and $200$ K, respectively.
	At $350$ K, the behavior of $F_\text{s}(q,t)$ curves resemble the behavior near the reflective wall. At $200$ K,  $F_\text{s}(q,t)$ curves show a different behavior, and long-time convergence of $F_\text{s}(q,t)$ curves (over a $q$ range) is observed.
	Qualitatively similar behavior is also observed near the graphene surface (\autoref{Fig:layer1} and \autoref{Fig:Fs-qs}). 
	As discussed above (the discussion of \autoref{Fig:layer1}) the long-time convergence of $F_\text{s}(q,t)$ curves originates from the large heterogeneity of particle displacements; therefore, it is correlated with the peak height of $\alpha_2^z(t)$.

	\section{Summary}

	In summary, we studied the 
	dynamics of a confined glass-forming molecular liquid perpendicular to the confining surface using molecular dynamics simulations with atomistic models.
	The model liquid was hydrogenated methyl methacrylate 
	(HMMA, C$_5$H$_{10}$O$_2$).
	For the confining surface, we examined different models, namely a periodic single layer of graphene and a frozen amorphous configuration of the bulk liquid (frozen wall). 
	The effect of the surface/liquid interaction strength was also studied by adjusting the interaction of the liquid with the graphene surface.  
	In the vicinity of graphene, the liquid forms a layered structure and HMMA molecules adopt parallel to surface alignment; however, the frozen wall does not alter the structure of the adjacent liquid.

	In the cases of favorable surface/liquid interaction, the molecules of the liquid are trapped in the vicinity of the surface. Because of this effect, the dynamics of the molecules located in the interfacial layer is dominated by the process of escaping of the molecules from the surface.
	The process of escaping from the surface was studied by measuring non-Guassian parameter ($\alpha_2^z(t)$), self-intermediate scattering function ($F_\text{s}(q,t)$), and mean-squared displacement ($\langle \Delta z ^2(t) \rangle$) of the liquid particles. 
	An attractive surface imposes strong heterogeneity on the displacements in the adjacent interfacial liquid layer; the heterogeneity is reflected in the large peak of the $\alpha_2^z(t)$ for the displacements of the particles of the interfacial layer.
	The long-time behavior of the $F_\text{s}(q,t)$ of the interfacial layer is dominated by the process of escaping from the surface, and over a wide range of $q$, long-time convergence of the $F_\text{s}(q,t)$ curves is observed.
	The $\langle \Delta z ^2(t) \rangle$ of the interfacial layer is also affected by the process of escaping from the surface. Specifically, when liquid particles start escaping the surface, $\langle \Delta z ^2(t) \rangle$ of the interfacial layer (when calculated relative to the initial location of the particles in the interfacial layer) exhibits an apparent supper-diffusive behavior.  
	
	A qualitative difference between the dynamics of the liquid near the graphene surface and near the frozen wall was observed. Near the graphene surface, a clear separation between the small-scale back-and-forth movement of the molecules in the vicinity of the surface and the process of escaping from the surface is observed. This separation is reflected in the three-step decay of $F_\text{s}(q,t)$ and three-step increase of $\langle \Delta z ^2(t) \rangle$.
	The first step comes from the rattling of the particles in the cages formed by their nearest neighbors; the second step is related to small-scale movements close to the surface, and the third step is related to the escaping of molecules from the surface. 
	However, a clear three-step relaxation is not observed in the vicinity of the frozen wall. 
	This difference probably originates from the parallel-to-surface orientation of the molecules in the vicinity of the graphene surface. 
	Because of the parallel orientation of the molecules, they can perform seesaw-like movements before escaping the surface.

	Finally, at high temperatures or in cases of negligible surface/liquid attraction strengths, 
	the liquid particles are not trapped close to the surface and the dynamical quantities of the interfacial liquid layer can be approximately described by the observables of a liquid in the vicinity of a reflective wall.

	\section{Supplementary Material}
	
	The following topics are discussed in Supplementary Material: 
	(1) Structure and dynamics of the bulk liquid. (2) A Fickian liquid near a reflective wall. (3) The layer-resolved $F_\text{s}(q,t)$ curves for HMMA/graphene system  at different temperatures. (4) Apparent super-diffusive regime of $\langle \Delta z^2(t) \rangle$. (5) $\alpha_2^z(t)$ of the interfacial layer  vs its $\langle \Delta z^2(t) \rangle$.

\end{document}


	
	\title{Supplementary Material:\\ Relaxation Dynamics of a Liquid in the Vicinity of an Attractive Surface: The Process of Escaping from the Surface}

\author{Alireza F. Behbahani}
\email{aforooza@uni-mainz.de}
\affiliation{Institut f\"{u}r Physik, Johannes Gutenberg-Universit\"{a}t Mainz, Staudingerweg 7, D-55099 Mainz, Germany}
\affiliation{Institute of Applied and Computational Mathematics, Foundation for Research and Technology - Hellas, Heraklion GR-71110, Greece}

\author{Vagelis Harmandaris}
\affiliation{Computation-based Science and Technology Research Center, The Cyprus Institute, Nicosia 2121, Cyprus}
\affiliation{Institute of Applied and Computational Mathematics, Foundation for Research and Technology - Hellas, Heraklion GR-71110, Greece}
\affiliation{Department of Mathematics and Applied Mathematics, University of Crete, Heraklion GR-71110, Greece}

	\date{\today}
	
	\begin{abstract}
		
	\end{abstract}
	
	\maketitle

\section{Structure and dynamics of the bulk liquid}	

\label{sec:bulk}

\begin{figure}[!htb]
	\centering
	
	\begin{subfigure}{0.4\textwidth} 
		\includegraphics[width=\textwidth]{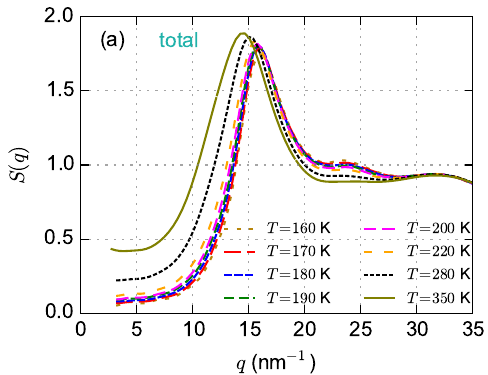}
	\end{subfigure}		
	\begin{subfigure}{0.4\textwidth} 
		\includegraphics[width=\textwidth]{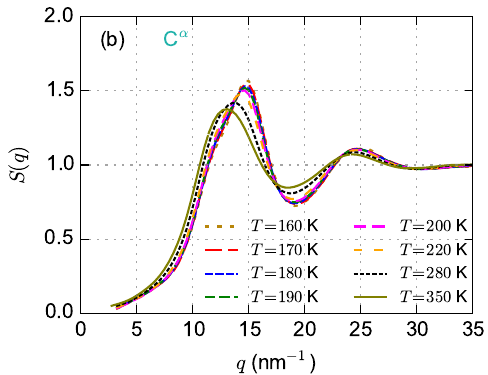}
	\end{subfigure}
	\caption{(a) Static structure factor of the HMMA liquid at different temperatures. (b) Partial structure factor of the $\alpha$-carbon atoms of HMMA at different temperatures.}
	\label{Fig:SM-sq} 
\end{figure}

The static structure factors of the model liquid, at different temperatures, are shown in \autoref{Fig:SM-sq}a.
\autoref{Fig:SM-sq}b shows the partial structure factor of the $\alpha$-carbon atoms of HMMA. 
With reducing temperature, the peak of $S(q)$ shifts to slightly larger $q$ values. This is the result of the increasing density of the liquid with decreasing temperature.  
The $F_\text{s}(q,t)$ curves are mainly calculated at $q = 14$ nm$^{-1}$ which is close to the peak position of the partial structure factor of the $\alpha$-carbon atoms.
Note that the experimental melting point of HMMA is $188$ K~\cite{kim2021pubchem}. However, in the simulation of the model HMMA liquid, we do not see any sign of crystallization or a time-dependent behavior at temperatures lower than $188$ K. This observation  originates from the much longer time scales of crystallization than the simulation time window. Furthermore, the melting temperature of the model HMMA liquid might deviate from that of the experimental sample (due to the choice of the force field).

\begin{figure}[!htb]
	\centering
	
	\begin{subfigure}{0.40\textwidth} 
		\includegraphics[width=\textwidth]{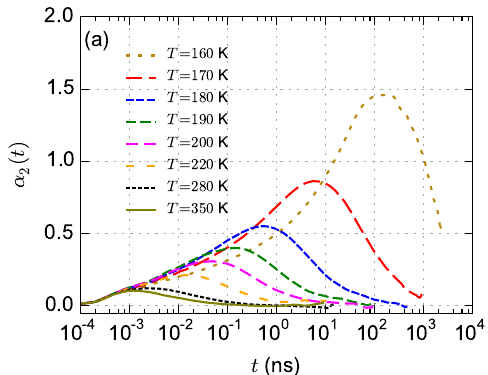}
	\end{subfigure}		
	
	\begin{subfigure}{0.40\textwidth} 
		\includegraphics[width=\textwidth]{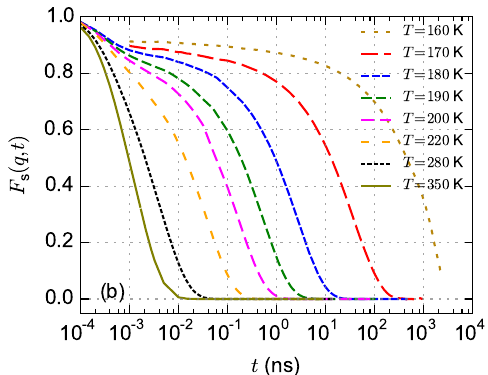}
	\end{subfigure}
	
	\begin{subfigure}{0.40\textwidth} 
		\includegraphics[width=\textwidth]{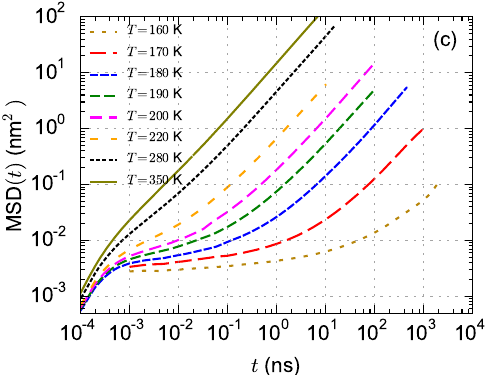}
	\end{subfigure}
	
	\caption{Dynamical observables for the model liquid at different temperatures. (a), (b), and (c) show non-Gaussian parameter, self-intermediate scattering function, and mean-squared displacement, respectively.     }
	\label{Fig:SM-bulk-dyn} 
	
\end{figure}

The dynamical observables for the bulk HMMA liquid at various temperatures are shown in \autoref{Fig:SM-bulk-dyn}.  
Upon reducing temperature, dynamical heterogeneities related to the cage effect increases. This is reflected in the higher peaks of $\alpha_2(t)$ at lower temperatures (\autoref{Fig:SM-bulk-dyn}a).
Furthermore, at lower temperatures, the cages are more rigid. This is reflected in the development of plateau regimes in the $F_\text{s}(q,t)$ and MSD$(t)$ curves (\autoref{Fig:SM-bulk-dyn}b-c).    	
The maximum of $\alpha_2(t)$ almost coincides with the onset of decay of 
$F_\text{s}(q,t)$ and also onset of increase of MSD$(t)$, after the cage regime.

\begin{figure}[!htb]
	\centering
	
	\begin{subfigure}{0.45\textwidth} 
		\includegraphics[width=\textwidth]{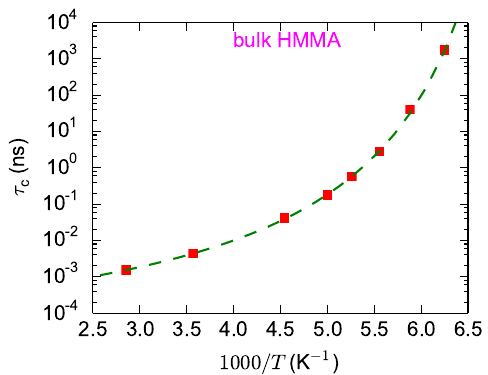}
	\end{subfigure}		
	
	\caption{Temperature dependence of the relaxation times of $\alpha$-process. The dashed curve is the fitted VFT curve on the data. }
	\label{Fig:SM-bulk-tau} 
	
\end{figure}

From the $F_\text{s}(q,t)$ curves we can calculate the relaxation times of the $\alpha$-process in the liquid. We fitted an stretched exponential function ($f(t) = A\exp(-(t/\tau)^\beta)$) on each $F_\text{s}(q,t)$ curve and calculated the relaxation times  through $\tau_c = \tau/\beta \Gamma(1/\beta)$, where $\Gamma$ is the Gamma function.      
The calculated relaxation times at different temperatures are presented in \autoref{Fig:SM-bulk-tau}. To describe the temperature dependence of $\alpha$-relaxation, we fitted a Vogel-Fulcher-Tammann (VFT) function ($\tau_c = \tau_\infty \exp(B/(T - T_0))$) on the relaxation times (see the dashed line in \autoref{Fig:SM-bulk-tau}). The fitting parameters are:
$T_0 = 129.0$ K, $B = 500$ K, and $\tau_\infty = 0.00016$ ns.

Usually, the \tg\ of a liquid is measured as the temperature at which the relaxation time of $\alpha$-relaxation equals around $100$ s. This time scale is very far from the accessible time scales of molecular simulations, however, from the fitted VFT relation we can calculate a rough estimate of \tg. 
Especially for the studied model liquid, a clear deviation from the Arrhenius temperature dependence of the relaxation times is seen in the temperature range of this study and hence the VFT parameters can be determined rather reliably.
For the model HMMA liquid, $T_\text{g} \approx 143$ K assuming $\tau_\text{c}(T_\text{g}) = 100$ s.
Hence, the temperature range shown in \autoref{Fig:SM-bulk-dyn}, approximately corresponds to $2.45$\tg\ down to $1.12$\tg.

\section{Dynamical observables for a Fickian liquid close to a reflective wall}

\begin{figure}[!htb]
	\centering
	\begin{subfigure}{0.45\textwidth} 
		\includegraphics[width=\textwidth]{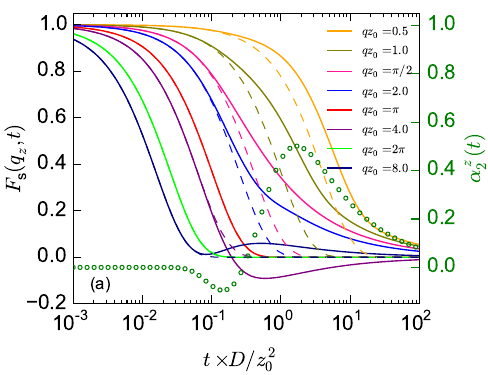}
	\end{subfigure}
	\begin{subfigure}{0.45\textwidth} 
		\includegraphics[width=\textwidth]{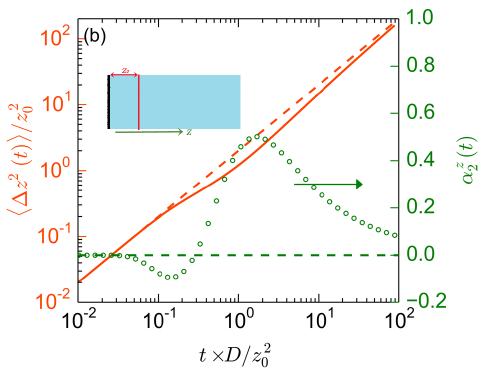}
	\end{subfigure}
	
	\caption{Solid curves and the curves with circle marker: the analytically  calculated dynamical observables for some tagged particles of a Fickian liquid  that at $t = 0$ are located at distance $z_0$ from a reflective wall (see inset of panel b). The dashed curves  display the observables for a bulk liquid with Fickian diffusion of particles.   }
	\label{Fig:SM-analytic-z0} 
\end{figure}

As mentioned in the main text, in bulk, if the particles have Fickian diffusion, the distribution of their displacements is Gaussian:
 $G_\text{s}(z,t) = \exp(-z^2/4Dt)/\sqrt{4\pi Dt}$,
 $\alpha_2^z(t) = 0$, $\langle \Delta z ^2(t) \rangle = 2Dt$, and $F_\text{s} (q_z,t) = \exp(-q_z^2 D t)$, where $D$ is the diffusion coefficient of the particles. 
Now, consider some tagged particles of a Fickian liquid that at $t = 0$ are located at a distance $z_0$ from an ideal planar reflective wall (that is, at $t = 0$ the spatial distribution of particles is $\delta(z-z_0)$, see the inset of \autoref{Fig:SM-analytic-z0}b). 
At the reflective wall $\partial c / \partial z = 0$, where $c$ is the concentration of the tagged particles.
The distribution of displacements of the tagged particles (displacements relative to the position at $t = 0$) can be calculated via reflection of the displacements at the wall. For these tagged particles: $G_\text{s}(z,t) = \big{[}\exp(-z^2/4Dt) + \exp(-(z+2z_0)^2/4Dt)\big{]}/\sqrt{4\pi Dt}, z\ge -z_0$.
With $G_\text{s}(z,t)$ at hand $F_\text{s} (q_z,t)$, $\langle \Delta z ^2(t) \rangle$, and $\alpha_2^z(t)$ can be calculated:
\begin{equation}
	\begin{aligned}	
		F_\text{s} (q_z,t) &=  = \int_{-\infty}^{\infty} G_\text{s}(z,t) \cos[q_z z] \mathrm{d}z 
		  \\
		\langle \Delta z ^2(t) \rangle &=  \int_{-\infty}^{\infty} G_\text{s}(z,t) z^2 \mathrm{d}z   \\
		\langle \Delta z ^4(t) \rangle &= \int_{-\infty}^{\infty} G_\text{s}(z,t) z^4 \mathrm{d}z   \\
		\alpha_2^z(t) &= \frac{1}{3} \frac{\langle \Delta z ^4 \rangle}{\langle \Delta z ^2 \rangle^2}-1\\
	\end{aligned}
	\label{Eq:reflective-wall}
\end{equation}

\begin{figure}[!htb]
	\centering
	
	\begin{subfigure}{0.45\textwidth} 
		\includegraphics[width=\textwidth]{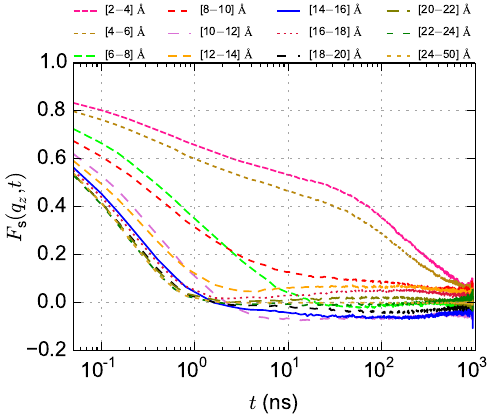}
	\end{subfigure}		
	
	\caption{$F_\text{s}(q,t)$ curves calculated for thin $2$ \AA\ layers in the confined liquid ($T = 200$ K). The $F_\text{s}(q,t)$ curves of some of the layers show a non-monotonic behavior which is consistent with the analytical results presented in \autoref{Fig:SM-analytic-z0}.   }
	\label{Fig:SM-Fs-2A} 
	
\end{figure}

We calculated the above integrals numerically. The results are presented in \autoref{Fig:SM-analytic-z0}. 
The geometric effect of the wall modifies the dynamical observables of the tagged particles.  
The behavior of $F_\text{s} (q_z,t)$ is interesting. Generally the $F_\text{s} (q_z,t)$ curves of the tagged particles differ from the bulk curves. However, at $q = n \pi/z_0$ ($n = 1, 2, 3, ..$), the $F_\text{s} (q_z,t)$ of the tagged particles are similar to the bulk ones.   
Furthermore, at some $q$ values (e.g., $q = 4/z_0$ and $q = 8/z_0$ in \autoref{Fig:SM-analytic-z0}), the  $F_\text{s} (q_z,t)$ curves are non-monotonic. For example, at $q = 4/z_0$, $F_\text{s} (q_z,t)$ becomes negative before converging to zero; or at $q = 8/z_0$, $F_\text{s} (q_z,t)$ becomes almost zero and then increases again. Such non-monotonic shapes can also be observed in the simulations through layer-resolved calculations of $F_\text{s} (q_z,t)$ for very thin layers. For example see \autoref{Fig:SM-Fs-2A}.

In the main text, we extend the above analysis to a layer of tagged particles that at $t = 0$ has a finite thickness of $d$  and is located in contact with a reflective wall.

\section{The layer-resolved $F_\text{s}(q,t)$ curves at different temperatures}

\begin{figure}[htb]
	\centering
	
	\begin{subfigure}{0.40\textwidth} 
		\includegraphics[width=\textwidth]{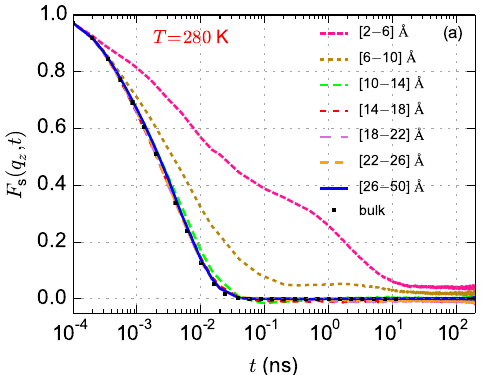}
	\end{subfigure}		
	
	\begin{subfigure}{0.40\textwidth} 
		\includegraphics[width=\textwidth]{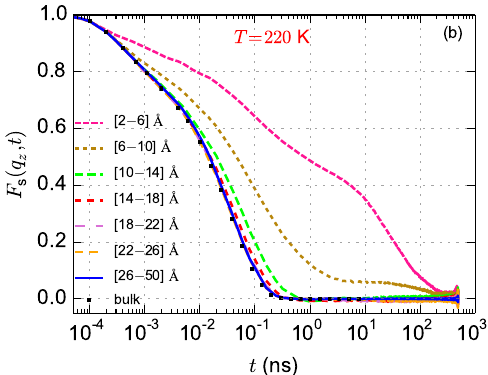}
	\end{subfigure}
	
	\begin{subfigure}{0.40\textwidth} 
		\includegraphics[width=\textwidth]{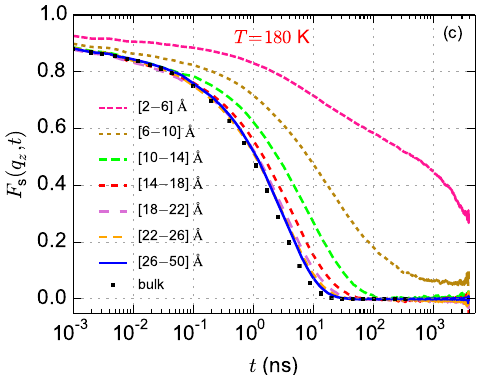}
	\end{subfigure}
	 
	\caption{Layer-resolved $F_\text{s}(q,t)$ curves for the confined liquid at (a) $T = 280$ K, (b) $T = 220$ K, and (c) $T = 180$ K ($q = 14$ nm$^{-1}$).}
	\label{Fig:SM-Fs-Ts} 
	
\end{figure}
 
\autoref{Fig:SM-Fs-Ts} shows the layer-resolved $F_\text{s}(q,t)$ functions of the confined liquid at different temperatures ($220$ K, $280$ K, and $180$ K). With reducing temperature, the length scale over which the confining surface modifies the dynamics of the liquid increases. At $T = 280$ K, the dynamics in the first and the second layers are affected by the confining surface. However, with reducing temperature, the dynamics in the further layers are also modified. At $T = 220$ K, the dynamical gradient propagates up to the fourth layer, and at $T = 180$ K, up to the fifth one. This trend correlates with the more pronounced layered structure of the liquid close to the surface at lower temperatures (see \autoref{Fig:SM-bulk-dyn}).
It is worth mentioning that the increase of the length scale of the surface effect is also observed near the frozen configuration of the bulk liquid, where the interfacial structure of the liquid is similar to the structure of the bulk liquid.

\section{About the apparent super-diffusive regime of $\langle \Delta z^2(t) \rangle$ }

\begin{figure}[htb]
	\centering
	
	\begin{subfigure}{0.45\textwidth} 
		\includegraphics[width=\textwidth]{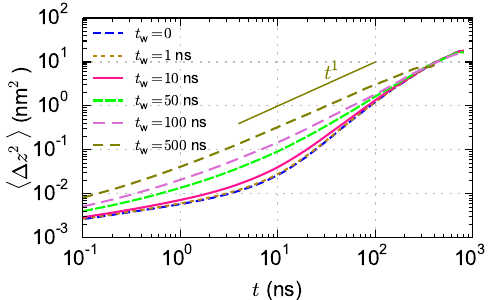}
	\end{subfigure}		
	 
	\caption{$\langle \Delta z ^2(t) \rangle$ for the particles of the first layer (particle present in the first layer at $t = 0$) calculated after different waiting times ($t_w$).}
	\label{Fig:msd-tw} 
	
\end{figure}

As shown in the main text (Figures 4c and 8c), the 
$\langle \Delta z ^2(t) \rangle$ of the first interfacial layer 
exhibits an apparent super-diffusive regime at long times, which is related to the escaping of molecules from the surface. However, as mentioned in the main text, 
this regime appears as a result of the calculation of the $\langle \Delta z ^2(t) \rangle$ relative to the initial position of the particles within the layer, and the articles do not have a real super-diffusive motion.  
To provide a support for this explanation, we measure 
$\langle \Delta z ^2(t) \rangle = \langle (z(t_\text{w} + t) - z(t_\text{w}))^2 \rangle$ for the particles of the first layer (particles that belong to first layer at time $t = 0$) calculated after different waiting times, $t_\text{w}$. 
In other words,  the layers are assigned at time $t =0$, however, 
$\langle \Delta z ^2(t) \rangle$ is calculated after a waiting time.
\autoref{Fig:msd-tw} shows the
result of this analysis.
As expected, by increasing the waiting time, the super-diffusive behavior gradually disappears. This shows that when some of the particles escape from the surface, they do not have super-diffusive behavior, however when displacements are calculated relative to the initial position of the particles in the vicinity of the surface, an apparent super-diffusive behavior is seen in the $\langle \Delta z ^2(t) \rangle$.

\section{$\alpha_2^z(t)$ of the interfacial layer  vs its $\langle \Delta z^2(t) \rangle$}

\begin{figure}[!htb]
	\centering
	
	\begin{subfigure}{0.45\textwidth} 
		\includegraphics[width=\textwidth]{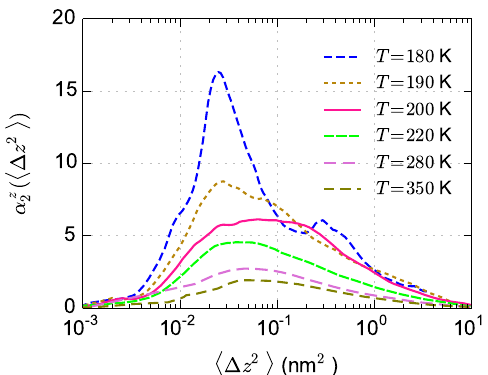}
	\end{subfigure}		
	
	\caption{The $\alpha_2^z(t)$ of the first layer (i.e., $[2-6]$ \AA\ from the surface) vs. its $\langle \Delta z^2(t) \rangle$. }
	\label{Fig:alpha-msd} 
\end{figure}

\autoref{Fig:alpha-msd} show $\alpha_2^z(t)$ of the first interfacial layer (i.e., $[2-6]$ \AA\ from the surface) vs its $\Delta z^2(t)$. In the temperature range studied here, the peak of $\alpha_2^z(t)$ appears around $2.5$--$4.3$ \AA$^2$.
With decreasing temperature, the position of the peak shifts to smaller values of 
$\Delta z^2(t)$.
These length scales correspond to the small-scale movements of the particles in the interfacial layer.